# Mars Obliquity History Constrained by Elliptic Crater Orientations

Samuel J. Holo (holo@uchicago.edu), Edwin S. Kite, and Stuart J. Robbins

Highlights:

- Obliquity and Mars crossers' inclinations control elliptic crater orientations
- Late Hesperian and younger craters possess a ~3:2 N-S orientation preference
- Inversion of Mars' obliquity history PDF reveals low ~3.5 Gyr mean obliquity

**Abstract.** The dynamics of Mars' obliquity are chaotic, and thus the historical ~3.5 Gyr obliquity probability density function (PDF) is highly uncertain and cannot be inferred from direct simulation alone. Obliquity is also a strong control on post-Noachian Martian climate, enhancing the potential for equatorial ice/snow melting and runoff at high obliquities (> 40°) and enhancing the potential for desiccating deep aquifers at low obliquities (< 25°). We developed a new technique using the orientations of elliptical craters to constrain the true late-Hesperian-onward obliquity PDF. To do so, we developed a forward model of the effect of obliquity on elliptic crater orientations using ensembles of simulated Mars impactors and ~ 3.5 Gyr-long Mars obliquity simulations. In our model, the inclinations and speeds of Mars crossing objects bias the preferred orientation of elliptic craters which are formed by low-angle impacts. Comparison of our simulation predictions with a validated database of elliptic crater orientations allowed us to invert for the best-fitting obliquity tracks. We found that since the onset of the late Hesperian, Mars' mean obliquity was likely low, between ~10° and ~30°, and the fraction of time spent at high obliquities > 40° was likely < 20%.

1. Introduction

Mars' obliquity, $\epsilon$, is currently ~25° but has changed dramatically over billions of years since solar system formation (Ward 1973, Touma & Wisdom 1993, Laskar 1993, Laskar 2004). The dynamics of Mars' obliquity are driven by secular spin-orbit resonances (Touma & Wisdom 1993, Laskar 1993). However, the obliquity evolution is sensitive to orbital properties that vary chaotically on timescales < 100 Myr (Touma & Wisdom 1993, Laskar 1993, Laskar 2004). Many geologic methods have been proposed to vault the fundamental barrier of the chaotic diffusion of the Solar System (e.g. Ma et al. 2017), but all are indirect. Further, only one-to-a-few transitions between low and high values of $\epsilon$ should occur (§3.1), preventing chaotic variations in obliquity from "averaging out" over billions of years. Thus, both the full obliquity history and the historical obliquity probability density function (PDF) are highly uncertain. Here we propose a direct method to constrain obliquity history.

Obliquity variations are a strong control on post-Noachian Martian climate (Jakosky and Carr 1985, Laskar et al. 2004). At low obliquities (< 25°), the Martian atmosphere is more likely to collapse at the poles (Kreslavsky and Head 2005, Phillips et al. 2011, Soto et al. 2015) and surface melting is unlikely (Fastook et al. 2012). At high obliquities (> 40°), models predict that water vapor pressure increases (Zent 2013), surface melting is more likely (Jakosky and Carr 1985), and strong dust storms initiate near the poles (Haberle et al. 2003). Insolation driven ice and snow melt has been proposed to explain observed sedimentary features near the equator (e.g. Kite et al. 2013, Irwin et al. 2015, Palucis et al. 2014). Further, low values of obliquity have been shown in models to dramatically enhance desiccation of deep aquifers via sublimation (Grimm et al. 2017).

To constrain the true effects of Mars' obliquity on post-Noachian climate, it is necessary to constrain the true obliquity history, and thus we sought a geologic constraint on Mars' obliquity history. Previous attempts to constrain Mars' obliquity from geologic features such as mid-latitude glaciers (e.g. Fassett et al. 2014) show that Mars' obliquity was high (~35°) for ~ 1 Gyr. However, no study has quantitatively constrained the full ~3.5 Gyr post-Noachian Mars obliquity PDF with geologic evidence. Here, we propose a new method of constraining the historical Mars obliquity PDF using the orientations of elliptic craters. The vast majority of craters on Mars are nearly circular, but impactors with small impact angles relative to the surface produce elliptic craters with major axes aligned with impactor velocity vector (Bottke et al. 2000, Collins et al. 2011). As a result, impactors that travel parallel to Mars' spin pole will create North-South oriented craters at the equator, and impactors that travel normal to the spin pole will create elliptic craters at all latitudes that are East-West oriented everywhere except near the pole (Figure 1). As the obliquity changes, the angles between impactors and the spin axis change, causing a change in predicted orientation of elliptic craters. We developed a numerical forward model of the effect of obliquity on the orientations of elliptic craters using realistic ensembles of simulated Martian impactor orbits and ~ 3.5 Gyr-long Martian obliquity simulations. We then used a validated version of a global database of Martian crater ellipticities and orientations (Robbins and Hynek 2012) and the ages of underlying geologic units (Tanaka et al. 2014) to invert for the true Martian obliquity history. From that we construct estimates of the mean obliquity and the number of years with $\epsilon > 40°$ (Figure 2).

## 2. Elliptic Crater Orientations Database

The Robbins global Mars crater database (Robbins and Hynek, 2012) contains measurements of crater ellipticities (ratio of major to minor axes lengths) and major axis orientations (absolute azimuth from due North) obtained from fitting ellipses to points traced around crater rims. The publicly available data has a known bug in these parameters where some major axis orientations have been shifted by 90°. While this error has been internally corrected (interested parties can obtain the corrected database from S.J.R.), there has been no independent check of the correction's accuracy. Thus, we carried out an exhaustive search for systematic inter-analyst variability (Appendix A) and found that, restricting our analysis to craters > 4km in diameter and degradation state ≥ 2 (i.e. filtering out the most degraded craters), inter-analyst residuals for both the ellipticity and orientation of craters are not systematic. Thus, we concluded that the Robbins database (Robbins and Hynek 2012) provides a suitable constraint on our model with no systematic inter-analyst error and well quantified random inter-analyst error.

Because our goal is ultimately to compare the population of elliptic crater orientations to predictions made by an ensemble of ~3.5 Gyr Mars obliquity simulations, we must filter out elliptic craters older than ~3.5 Gyr. Individual crater ages are difficult to constrain, but maximum crater ages are constrained by the age of the underlying geologic unit. Thus, we can use the Tanaka et al. (2014) global geologic map of Mars to identify the maximum ages of craters. Because not all of the geologic localities have well determined individual ages due to small surface areas, we rely on the reported geologic epoch of each locality. All terrains listed as Amazonian, early/middle/late Amazonian, Late Hesperian, and Amazonian-Hesperian were included in our analysis (~45% of the Martian surface area). The maximum ages associated with each epoch are taken from Tanaka et al. (2014) assuming that the Amazonian-Hesperian units have a maximum age halfway through the late Hesperian. We bracket our uncertainty in absolute ages by using both the Hartmann and Neukum chronology ages as reported in Tanaka et al. (2014).

After filtering for maximum geologic ages, minimum diameter, and minimum preservation state, we are left with N = 1502 elliptic craters (crater properties listed in Supplementary Data Table 1). Roughly 2/3 of the craters are found in the northern plains, while the remaining craters are found in the Tharsis region and other young volcanic and apron units in the Southern hemisphere (Figure 3, Tanaka et al. 2014). The craters possess a preference for North-South oriented major axes that varies in strength with crater diameter and latitude (Figure 3). Integrated over diameter and latitude, the global preference for North-South vs. East-West oriented elliptic craters is roughly 3:2 (Figure 4).

To estimate our uncertainty in the major axis orientation PDF, we cannot use the typical $\sqrt{N}$ estimates. This is because elliptic crater locations and the observed orientation distribution are a function of both impact processes and geologic masking due to erosion and resurfacing. Thus, the errors in our estimate are not necessarily random and are drawn from a distribution that is not well-described by a Poisson process. To overcome this barrier, we instead assume that the observed crater orientations represent a sample from some distribution with unknown shape. In this framework, we can take the observed crater orientations as a non-parametric estimate of the underlying distribution from which they are drawn and perform a procedure called bootstrapping (e.g. Robbins et al. 2018). The bootstrapping procedure involves resampling a population (usually taking the same number of samples as is in the original dataset) with replacement such that some values can be chosen multiple times. We computed 1000 bootstrapped re-samples of the data, and computed a kernel density estimate (bandwidth of 5° as measured in the data-validation process) of the resulting orientations. The envelope of possible orientation PDF's is shown in Figure 4.

3. Forward Model Description

We developed a forward model for the PDF of elliptic crater orientations on Mars that contains two major components: 1. An ensemble of possible Mars obliquity histories (Kite et al. 2015) and 2. a long-term cratering model that, using a forward N-body simulation, estimates the locations, sizes and orientations of elliptic impact craters as a function of obliquity. From this we can determine which obliquity histories are most likely to reproduce the observed elliptic crater orientations found in the now-vetted Robbins database (Robbins and Hynek, 2012).

*3.1 Mars Obliquity History Ensemble*

We generated an ensemble of forward integrations of Mars' obliquity to estimate the range of ~3.5 Gyr Mars obliquity PDF's (Kite et al. 2015). To do so, we computed 38 orbital trajectories for Mars using the *mercury6* N-Body code (Chambers 1999). For each of these trajectories, the present day solar system with a small randomly generated offset in the position of Mars was taken as the initial condition. Each of these orbital trajectories was used to drive 24 instances of an obliquity code (Armstrong et al. 2014) where the initial obliquity was drawn from the PDF's of 3 Ga Martian obliquity reported by Laskar et al. (2004). Mars' obliquity undergoes only one-to-a few transitions between high and low values in ~3.5 Gyr, which is consistent with results for a moonless Earth (Lissauer et al. 2012, Li & Batygin 2014). Further, Mars' obliquity exhibits ~15° peak-to-trough amplitude quasi-periodic non-chaotic oscillations on Myr timescales (Laskar et al. 2004). Thus, we restricted the 38 x 24 = 912 obliquity time series (with output every 1000 years) to the 250 that were stable (i.e. Mercury does not fall into the sun and the model run did not crash) for > 3.61 Gyr and had a final Myr mean obliquity between 10° and 40° (i.e. close to the modern value). This ensemble acts as a random sample from a large chaotic phase space of possible obliquity tracks. While none of these integrations is the true obliquity history, they represent a realistic distribution of obliquity PDF summary statistics. Thus, we used these 250 obliquity tracks to calculate estimates of the prior distribution (i.e. before considering the elliptic crater constraint) of the historical Martian mean obliquity and time spent with $\epsilon > 40°$.

*3.2 Cratering Model*

In reality, Martian impactors come from all directions and with varying speeds. To estimate the Gyr-averaged impactor population on Mars, we assumed that the main asteroid populations are in steady state (Bottke et al. 2002) and that asteroids dominate the crater record (Bottke et al. 2000). With these assumptions, modern solar system impactors can be taken as representative of solar system impactors in the last few Gyr. We initiated a 10 Myr modern solar system integration with the modern Mars crossing objects (as reported by the Minor Planet Center in September 2016) as test particles and the planets as massive bodies. Mars crossers were defined as having perihelion distance less than $a \times (1 + e)$ and aphelion distance greater than $a \times (1 - e)$, where $a = 1.52$AU is the semi-major axis of Mars and $e = 0.15$ was chosen to include the entire range of Mars eccentricities observed in our simulations (Figure 5). Instances of an object passing within one Hill radius of Mars were taken as a close encounter, and the relative velocity of each close encounter asteroid relative to Mars was stored. From this we calculated the encounter speed, $v_\infty$, (magnitude of the relative velocity vector) and encounter inclination, $i_\infty$, (angle between relative velocity vector and plane of Mars' orbit). Note that the effect of relative velocity causes objects to possess encounter inclinations much larger than their orbital inclinations. After a single close encounter, the asteroid was removed from the simulation (we confirmed that the total population of Mars crossers does not change significantly over the duration of our simulation). To ensure that our ensemble of close encounters is statistically complete and does not possess an inclination bias, we considered only bright objects with absolute magnitude H ≤ 14 (Bottke et al. 2002). Following the completeness check of JeongAhn and Malhotra (2015), we confirmed that the absolute magnitude distributions for high and low inclination objects (with an orbital inclination cutoff of 15°) follow a power law for H ≤ 14 and are indistinguishable via a two sample Kolmogorov-Smirnov test (p = 0.05). This procedure resulted in 124 total close encounters.

From this ensemble, we seeded a forward model (Figure 6) of crater locations, sizes, and orientations as a function of obliquity. We assume the impactor follows an angular-momentum-conserving hyperbolic orbit about the center of Mars (LeFeuvre and Wieczorek 2008), and thus, for a given $\{i_\infty, v_\infty\}$, all impactors within a circular gravitational cross section with radius $\tau$ must impact the surface (Figure 6). To generate statistics on the long term predicted distribution of orientations as a function of location and size, we generated $10^7$ random samples from the encounter $\{i_\infty, v_\infty\}$ distribution obtained from the N-body simulations. For each sampled inclination-speed pair, we generate a random trajectory from a uniform distribution within the gravitational cross section (i.e. the square of the impact parameter $b^2$ is distributed uniformly in $[0, \tau^2]$, and the impact argument δ is distributed uniformly in $[0,2\pi]$) and analytically solve for the velocity and position of the impactor at the time of impact, assuming that the spin pole precession angle (i.e. the angle between the axes of rotation for encounter inclination and obliquity in the model geometry) is uniformly distributed in $[0,2\pi]$ (LeFeuvre and Wieczorek 2008).

From this forward model we produced two {latitude, azimuth, impact velocity, obliquity, impact angle} ensembles: one with obliquity 0° and all impact angles (i.e. including circular craters), the other varying obliquity between 0° and 90° for impact angles less than the critical angle for elliptic crater formation (calculated using the scaling relation of Collins et al. 2011 that reports the critical impact angle as a function of cratering efficiency). We compared the first ensemble with the observed size frequency distribution (SFD) of craters between 4 and 32 km in the Robbins database (Robbins and Hynek 2012) and tuned a single power law SFD for the impactors that, via a realistic crater diameter scaling (eq. A7 in Collins et al. 2011, in which crater diameter depends positively on impactor speed), provides a best fit (minimized $\chi^2$ statistic for $\sqrt{2}$ scaled bin widths results in power law slope of 1.5) to the observed SFD when convolved with the impact conditions. Our model fit and the data were found to be in close agreement. We then applied the tuned impactor SFD and crater diameter to the elliptic crater ensemble and produced a {latitude, azimuth, diameter, obliquity} ensemble. The ensembles shift from a North-South orientation preference at low obliquities to an East-West preference at high obliquities (Figure 7). This is largely due to the fact that, while there are roughly twice as many low-inclination close encounters as high-inclination ones, the high inclination objects have larger encounter speeds and thus contribute more to the ≥ 4 km crater population (Collins et al. 2011).

4. Model-Data Comparison

To compare our model outputs with the ~1500 observed elliptic craters, we must account for several factors. First, for a particular 3.5 Gyr obliquity PDF, the possible obliquities during the time of a crater's emplacement are constrained by that crater's maximum age. Second, for a given obliquity, our forward model ensemble represents the orientation preference of elliptic craters once the surface is essentially saturated with elliptic craters (recall there are ~500,000 elliptic craters in each single-obliquity ensemble). However, it is possible in principle that an obliquity PDF that produces an E-W preference after saturating the surface with elliptic craters could produce a N-S preference when only 1500 craters are selected. Finally, the orientation preference of elliptic craters is not independent of latitude or diameter, but, as discussed above, the latitude-diameter distribution of elliptic craters is largely a function of geologic resurfacing processes and does not have well-parameterized errors.

To overcome these difficulties, we developed a stochastic fitting algorithm that determines which obliquity PDF's are most likely to reproduce the observed orientation preference given the selection of ~1500 elliptic craters at the observed latitudes and diameters. For a particular obliquity PDF, we take a bootstrapped re-sampling of the observed elliptic craters with the same number of craters as the original sample and assign random time stamps to each crater between the present and each crater's maximum age (from the geologic unit it lies on). This procedure assumes that the flux of craters has been constant between the late Hesperian and the present. We then, for each crater, randomly choose a crater from the model {latitude, diameter, obliquity, azimuth} ensemble with the constraints that the obliquity corresponds to the obliquity at the assigned time stamp, that the latitude is within $2.5^o$ of the observed crater's latitude, and that the diameter is within 5% of the observed crater's diameter. This has the effect of both incorporating the maximum age constraint and approximately fixing the predicted latitude-diameter distribution to the observed distribution. Thus, the orientations from the ensemble-sampled model craters represent a crater orientation prediction for the chosen obliquity PDF.

We can then calculate a goodness-of-fit by calculating the two-sample Kolmogorov Smirnov statistic between the bootstrapped re-sample of observed crater orientations and the predicted crater orientations. We repeat this process for each of our ensemble obliquity PDF's and determine which provides the best goodness-of-fit. This process is highly stochastic, and, as expected, obliquity PDF's which frequently produce poor fits can occasionally produce good fits. Thus, we repeat the entire process 1000 times and determine with what frequency each obliquity

PDF provides the best fit (Figure 8). We can then weight each obliquity PDF by its relative best-fitting frequency to produce weighted PDF's of both the mean obliquity and the fraction of time spent since the onset of the late Hesperian with obliquity > 40° (Figure 9). We repeat this entire analysis with both the Hartmann and Neukum chronology ages (as reported in Tanaka et al. 2014) to account for our uncertainty in absolute ages.

We found that Mars' mean obliquity from late Hesperian onwards was likely low, between ~10° and ~30°, and we reject at the 95% confidence level that mean obliquity was greater than ~33° (Figure 9). Applying the elliptic crater orientation constraint shifted this 95th percentile upper bound from ~44° in our unweighted mean obliquity PDF. Further, we found that the fraction of time spent since the onset of the late Hesperian with obliquity > 40° was also likely low, < ~20%, and we reject at the 95% confidence level that this fraction was > ~40%. This 95th percentile upper bound shifted from ~70% in our unweighted PDF of fraction of time spent at high obliquities. Our inverted distributions for both mean obliquity and fraction of time spent at high obliquities indicate that the true values are likely lower than the central expectations from our obliquity history ensemble. We failed to reject low obliquity cases that have means of ~10° and that spend < ~10% of time at high obliquity.

5. Discussion

The model relies on the assumption that our simulated encounter inclination-speed ensemble is representative of ~3.5 Gyr of Martian history and is unbiased. Increasing the relative abundance or speed of high inclination impactors biases the model to higher obliquity values. Conversely, increasing the abundance or speeds of low inclination impactors shifts the model output to lower obliquity values. However, the Martian impactor flux has likely been stable for over 3 Gyr (Nesvorny et al. 2017 and Robbins 2014). In addition, we took a conservative cutoff in absolute magnitude of 14 for our impactors, which is sufficient for preventing biases in the absence/presence of modern high inclination impactors (Bottke et al. 2002, JeongAhn and Malhotra 2015). An additional source of potential inclination bias in our ensemble arises from the inclusion of high-inclination objects from the Hungaria region, the stability of which (and thus the ~3 Gyr history of which) is still debated (Cuk and Nesvorny 2017, Cuk 2012, Bottke et al. 2012). However, of the close encounters, only one was an asteroid sourced from the Hungaria region.

The model is also sensitive to the impactor SFD, as crater diameter scales with impactor diameter (Collins et al. 2011). Although the absolute magnitude cumulative distribution functions for high and low impactors are indistinguishable, there is a potential for an albedo bias in inclination that is offset by a bias in impactor size. We performed a sensitivity test and found that a 10% bias in albedo for high inclination objects biased the model by the equivalent of just a few degrees in obliquity. We concluded that the model can robustly reject high or low obliquity solutions even in the presence of a small albedo bias. Thus, while we suspect any biases that arise from sampling Hungaria derived objects or correlations between albedo, inclination, and speed are minor, this is an assumption that must be interpreted as a caveat for our work.

Uncertainty in the crater database is also responsible for uncertainty in the model inversions. In particular, the small number of craters (~1500) is responsible for uncertainties in the true strength of the North-South orientation preference (Figure 4). Inclusion of craters with much smaller diameters (soon possible with CTX imagery) would tighten the uncertainties in the true azimuth PDF and allow the model to more precisely constrain properties of the true obliquity history. We note that while we verified that inter-analyst error in ellipticity and azimuth measurements can be attributed to a random process, we did not account for other systematic errors that would affect all analysts (e.g. those that could be attributed to second order projection effects or lighting angles).

Geologic surface processes and crater collapse are in principle capable of modifying crater morphologies (e.g. Weiss and Head 2015). However, it is unclear whether or not these processes actually produce elliptic craters with a systematic orientation bias. We repeated our analysis with a minimum ellipticity cut-off of 1.2, to test whether our conclusions change with the more conservative dataset. Overall, the results are consistent with our original analysis, supporting a low mean obliquity and < 40% of the late Hesperian onward time spent at high obliquity (Supplementary Figure 1), however, the $95^{th}$ percentile upper bounds shift to ~$40^o$ and ~55% for mean obliquity and high obliquity time fraction, respectively. We attribute these weaker upper bounds to the fact that only 276 craters meet the 1.2 ellipticity cutoff. As a result, our model-data comparison has a harder time distinguishing between acceptable obliquity PDF's (Supplementary Figure 2). Further, it is possible in principle that a thicker atmosphere in the late Hesperian may have biased the population of small elliptic craters, but we note that the global fraction of craters that are elliptic is the same for craters > 4 km in diameter and for craters > 10 km in diameter, which suggests that any such effect is negligible.

The model is unable to distinguish obliquity tracks that start low and end high from those that exhibit the opposite behavior. This is largely due to the small number of craters in Amazonian terrains, which possess an azimuth distribution indistinguishable from that of craters on late Hesperian units. In addition, the model selects different obliquity PDF's in the Hartmann and Neukum cases, but the systematic rejection of high obliquity solutions is common to both of them. This implies that our estimated upper bounds on the time spent above 40° and the mean obliquity are robust to uncertainties in age estimates from crater counts. However, we cannot say how much of the total time with $\epsilon > 40°$ occurred in the Amazonian, as opposed to the late Hesperian.

In theory, the model could be applied to understanding the elliptic crater populations from earlier in Mars' history and also on other planetary bodies. However, application to early Mars history would require verification of ellipticity and orientation measurements of more degraded craters on Noachian terrain. In principle, models of late heavy bombardment impactors (e.g. Nesvorny et al. 2017) could be tested by using Noachian crater orientations. Elliptic crater analysis also has the potential to test hypotheses of true polar wander early in Mars' history. For example, one could forward-model potential changes in true pole and compare the Noachian and post-Noachian crater orientations. We ignored any effects of true polar wander in our model as there is no definitive evidence of significant (more than a few degrees) true polar wander in the late Hesperian or onward (Kite et al. 2009, Matsuyama and Manga 2010). However, late stage true polar wander by more than a few degrees could affect the results presented here should definitive evidence for it be found.

6. Conclusions

We developed a forward model of the effect of obliquity on the distribution of elliptic crater major axis orientations. Using the model alongside the measured elliptic crater orientation distributions and an ensemble of obliquity histories, we inverted an estimate of the ~3.5 Gyr Martian obliquity PDF. We produced weighted estimates of the mean obliquity and compared them to unweighted estimates from our prior distribution of obliquity PDFs. Applying the geologic constraint decreased the most probable mean obliquity from ~33° to between ~10° and ~30° and moved the 2-$\sigma$ upper bound on mean obliquity from ~44° to ~33°. We also found that applying the geologic constraint reduces our weighted estimate of fraction of time spent above 40° obliquity from ~35% to < ~20% and moved the 2-$\sigma$ upper bound from ~70% to ~40%. The exact values of these estimates are sensitive to the uncertainties in absolute dates of geologic units as well as deviations from a constant crater flux. However, they robustly decrease both the estimates of, and the upper bound on, the fraction of time after the onset of the late Hesperian spent with obliquity > 40°. These results are consistent with findings from both Laskar (2004) and Bills (2006). This updated upper bound may be useful when calculating the cumulative effects of obliquity on both insolation driven snow/ice melting (Kite et al. 2013, Irwin et al. 2015, Palucis et al. 2014) and rapid loss of atmospheric water (Grimm et al. 2017).

Acknowledgments: We thank Bill Bottke and H.J. Melosh for helpful reviews, and Bill McKinnon for editorial handling. We thank Dan Fabrycky, Dorian Abbot and Susan Kidwell for helpful discussions. We thank David Nesvorny for sharing data. David Mayer assisted with GIS and high-performance computing. Computing resources were provided by the Research Computing Center at the University of Chicago. We acknowledge funding from NASA (NNX16AG55G).

Appendix A: Crater Ellipticity and Orientation Measurement Validation

To validate the corrected Robbins database (Robbins and Hynek, 2012), we examined de-projected measurements to search for and identify any systematic inter-analyst error. To do so, we divided the Robbins database into two dimensional ellipticity-diameter bins, where ellipticity bins had a minimum value of 1.1 and width 0.1 and diameter bins had a minimum value of 5 km and √2 scaled widths. We randomly sampled up to 10 craters from each two dimensional ellipticity-diameter bin in the updated database (some bins contained fewer than 10 craters). We compiled a set of 563 craters spanning a wide range of ellipticities, diameters, degradation states, and latitudes. We independently retraced these craters, removed first order projection effects, and fit ellipses to them using a direct non-linear least squares procedure (following Fitzgibbon et al., 1999 and Robbins and Hynek, 2012).

We compared our measured ellipticities and orientations to those measured by Robbins and Hynek (2012) and reported in the correction. The inter-analyst residuals (defined as the difference between the values measured by Robbins and Hynek (2012) and our re-measured values for each crater) had both a non-zero mean and a non-zero skewness, even after filtering out the most degraded class of craters (degradation state 1 in Robbins and Hynek, 2012). To assess whether these residuals can be attributed to random error, we resampled the residuals with replacement 10,000 times to produce a bootstrapped ensemble of equally likely residual distributions. For each resampled population, we calculated the residual means and skewnesses for both ellipticity and major-axis orientation. Histograms of the bootstrapped means and skewnesses show that the inter-analyst residuals for both ellipticity and orientation have means and skewnesses that are not significantly different than zero (Supplementary Figure 3). We concluded that for all craters other than the most degraded class, measurements of ellipticity and major-axis orientation show no systematic inter-analyst error.

Restricting this analysis to craters with modest ellipticities < 1.3 (as this is the region in which most of our final data lies), we found that orientation residual means, orientation residual skewness, and ellipticity residual skewness are still not significantly different than 0. While the ellipticity residuals showed a slight positive skewness, this only affects the acceptance/rejection of a crater for our analysis and does not systematically alter our orientation PDF. For craters with modest ellipticities, we found that the orientation residuals are roughly normally distributed with a standard deviation of 5°. Thus, we conclude that the Robbins database (Robbins and Hynek 2012) provides a suitable constraint on our model with no systematic inter-analyst error and well quantified random inter-analyst error.

Appendix B: Forward Model Validation

We validated our forward model by injecting artificially simple impactor populations and confirming that the resultant elliptic crater orientations and latitudes were consistent with the expected outcome. In particular, we injected impactors only from the orbit plane and varied the obliquity. At 0° obliquity, all elliptic craters were perfectly E-W oriented and located at the equator. At 90° obliquity, all elliptic craters were perfectly N-S oriented and distributed uniformly in latitude. We performed analogous tests with impactors coming from a plane inclined 90° with respect to the orbit plane and found consistent results. We also injected impactors with isotropic velocity vectors and found that the resultant orientations were uniformly distributed and invariant with obliquity. Further, in the isotropic case the number of craters dropped off sinusoidally with latitude (uniform crater density in latitude), independent of obliquity choice. We also confirmed that the effects of gravity focusing were properly accounted for by both visual inspection of impactor velocity vectors and checking that impactor angular momentum relative to Mars is conserved. Finally, we confirmed that the fraction of craters in our model, ~5%, that are elliptic is consistent with the observed fraction in the Robbins and Hynek (2012) database.

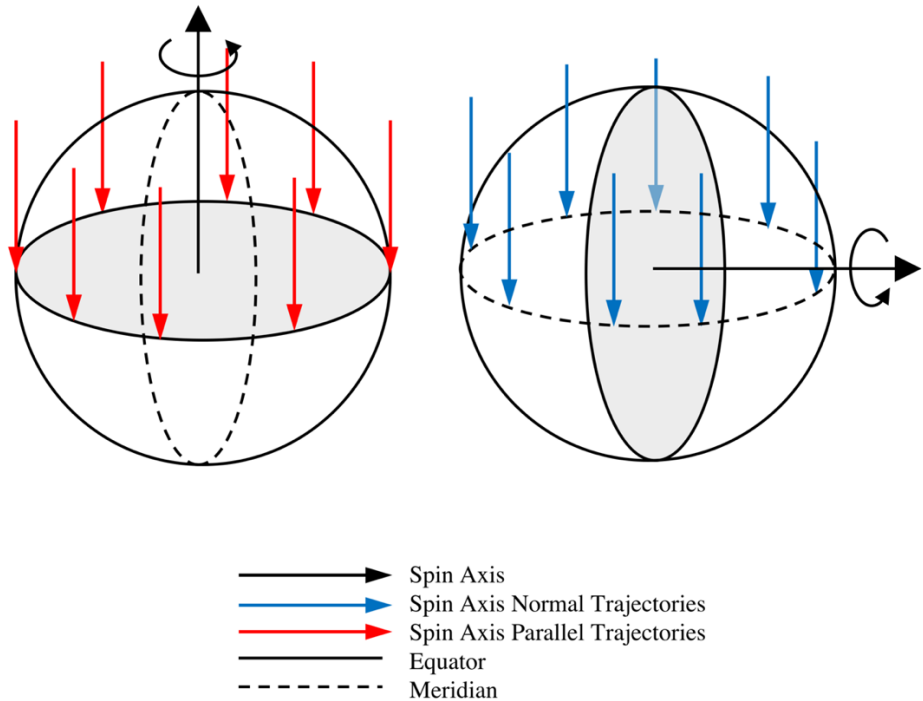

**Figure 1.** Schematic illustrating the basic principle in our model. Spin axis parallel impactors create N-S elliptic craters near the equator, while spin axis normal impactors produce elliptic craters that are E-W oriented at all latitudes except near the pole. The effects of gravity focusing are not shown here for simplicity.

**Note to Editor: High resolution file has been attached with submission. This figure has been formatted to be 1.5 column widths.**

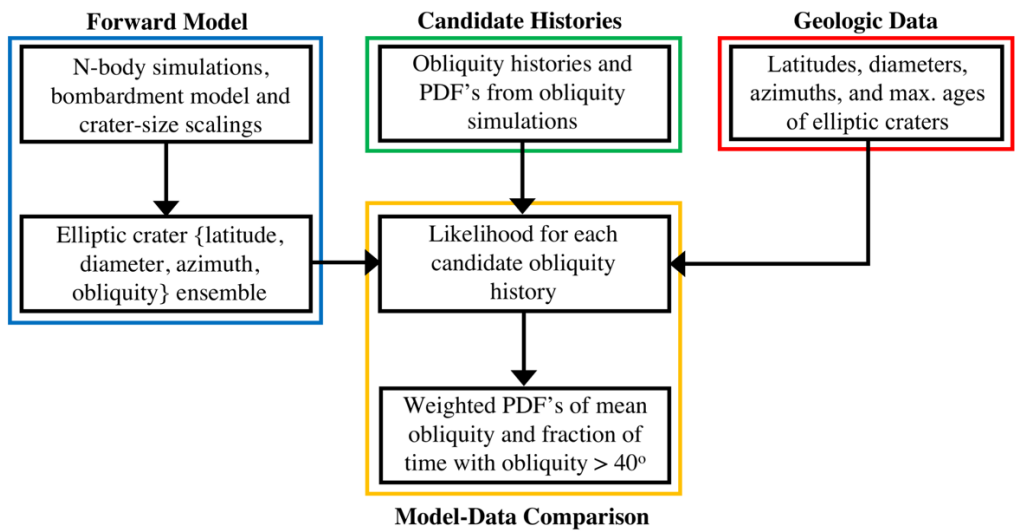

**Figure 2.** Schematic showing the model workflow.

**Note to Editor: High resolution file has been attached with submission. This figure has been formatted to be 1.5 column widths.**

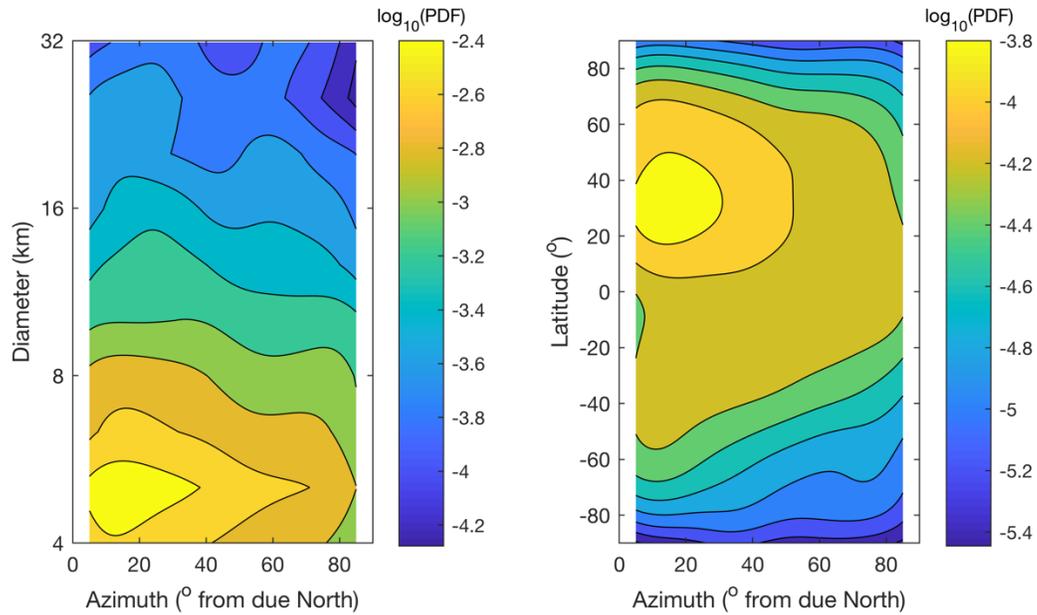

**Figure 3.** Smoothed heat maps of crater diameter vs. major axis orientation (left) and latitude vs major axis orientation (right). Azimuth data has been trimmed below 5° and above 85° to eliminate artifacts of the smoothing kernel (bandwidth of 5° in azimuth). Diameters were smoothed in $\log_{10}$ space (bandwidth of .05).

**Note to Editor: High resolution file has been attached with submission. This figure has been formatted to be 2 column widths.**

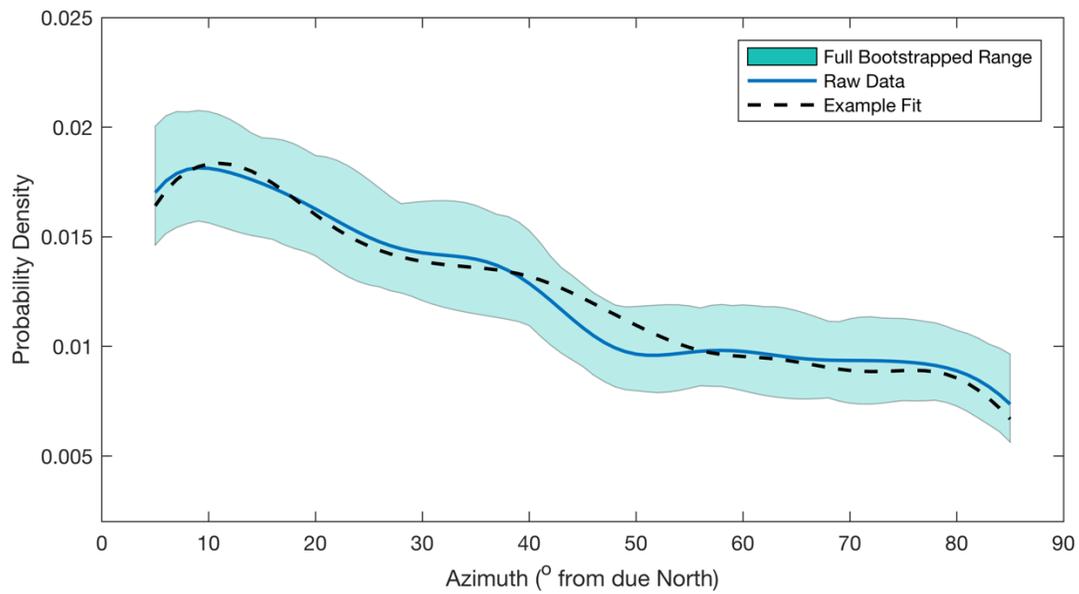

**Figure 4.** Smoothed Gaussian kernel estimates of crater azimuth probability density (bandwidth of 5°). Azimuth data has been trimmed below 5° and above 85° to eliminate artifacts of the smoothing kernel. Example fit was chosen from the stochastic fitting scheme (see Section 4).

**Note to Editor: High resolution file has been attached with submission. This figure has been formatted to be 2 column widths.**

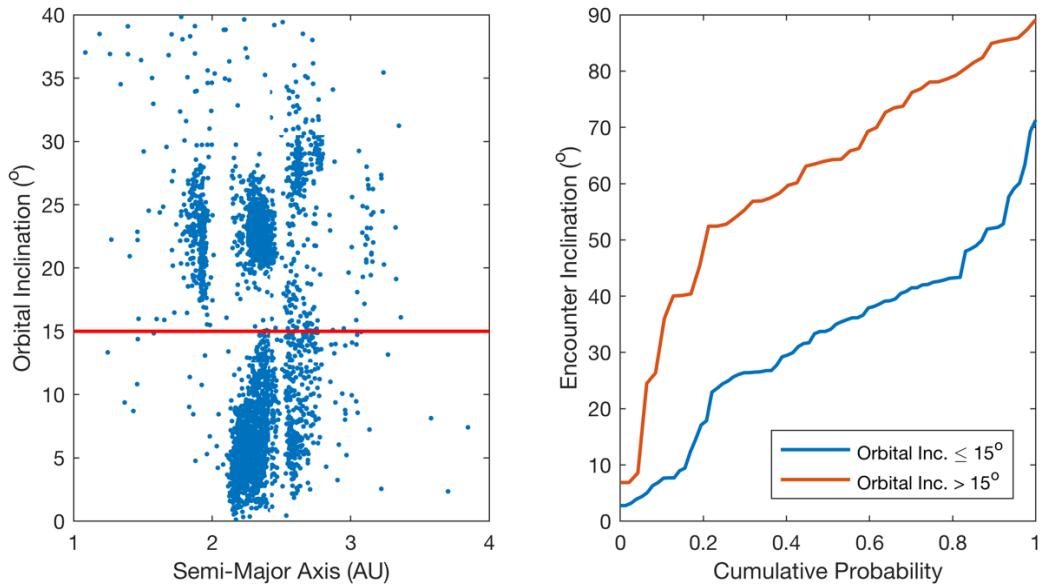

**Figure 5.** Left: Scatter plot of *orbital* inclination vs. semi-major axis of potential Mars crossing objects. The red line separates what we refer to as high and low inclination objects. For absolute magnitude < 16, there are 3282 objects. Right: cumulative probability function of the *encounter* inclinations, upon entering Mars' Hill sphere, of the 124 objects that experience a close encounter.

**Note to Editor: High resolution file has been attached with submission. This figure has been formatted to be 2 column widths.**

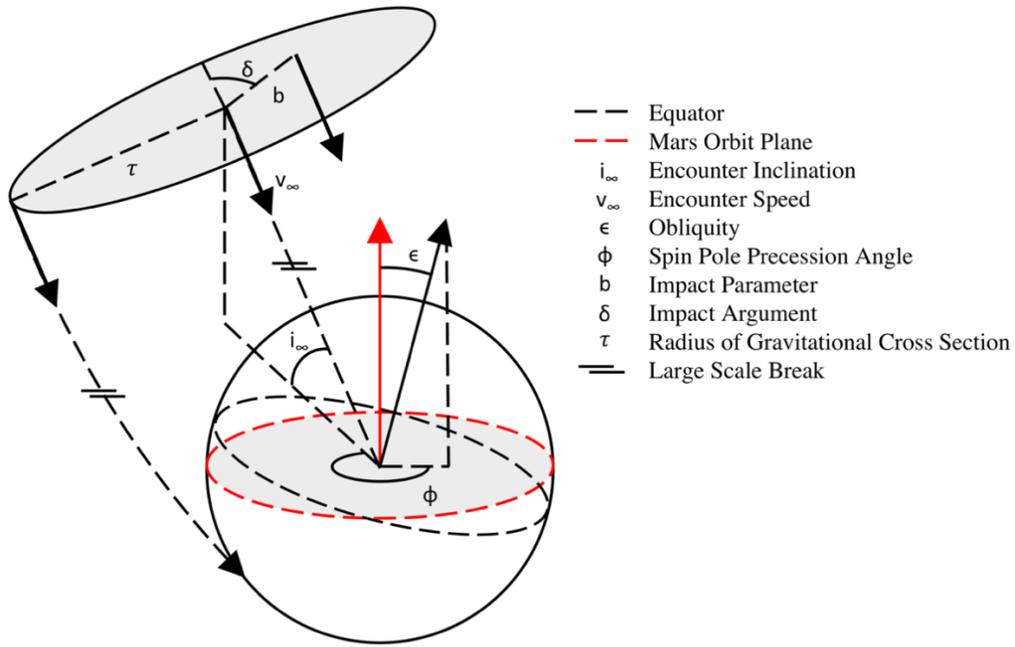

**Figure 6.** Schematic illustrating the geometry of the impact model.

**Note to Editor: High resolution file has been attached with submission. This figure has been formatted to be 1.5 column widths.**

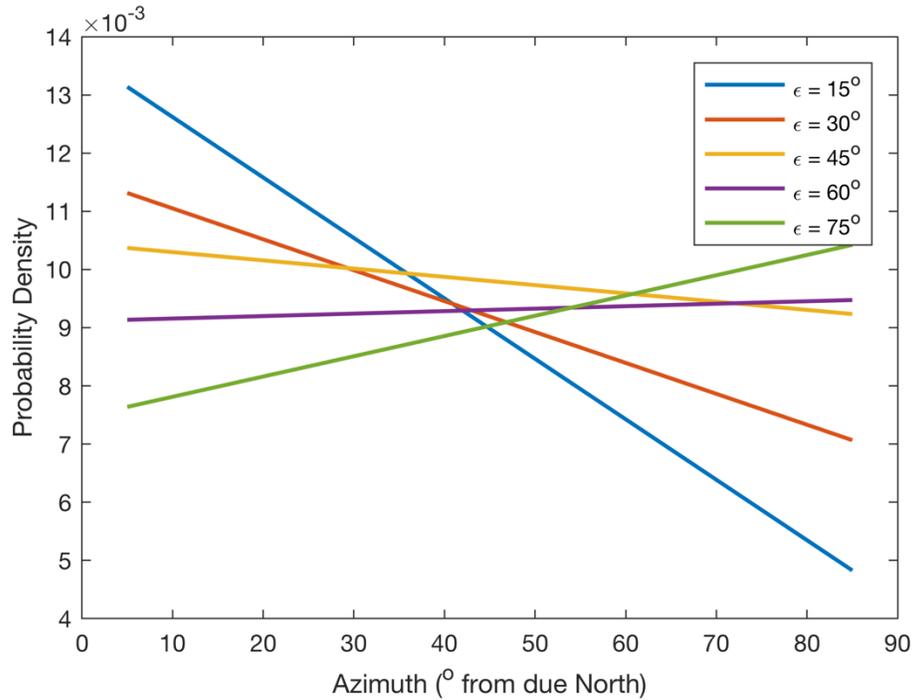

**Figure 7.** Gaussian kernel estimate of crater azimuth PDF (5° bandwidth) as a function of a single fixed obliquity prior to geologic correction. At low obliquities, there is a preference for North-South oriented elliptic craters. This trend is reversed at high obliquities. Azimuth data has been trimmed below 5° and above 85° to remove artifacts of the kernel smoothing process.

**Note to Editor: High resolution file has been attached with submission. This figure has been formatted to be 1.5 column widths.**

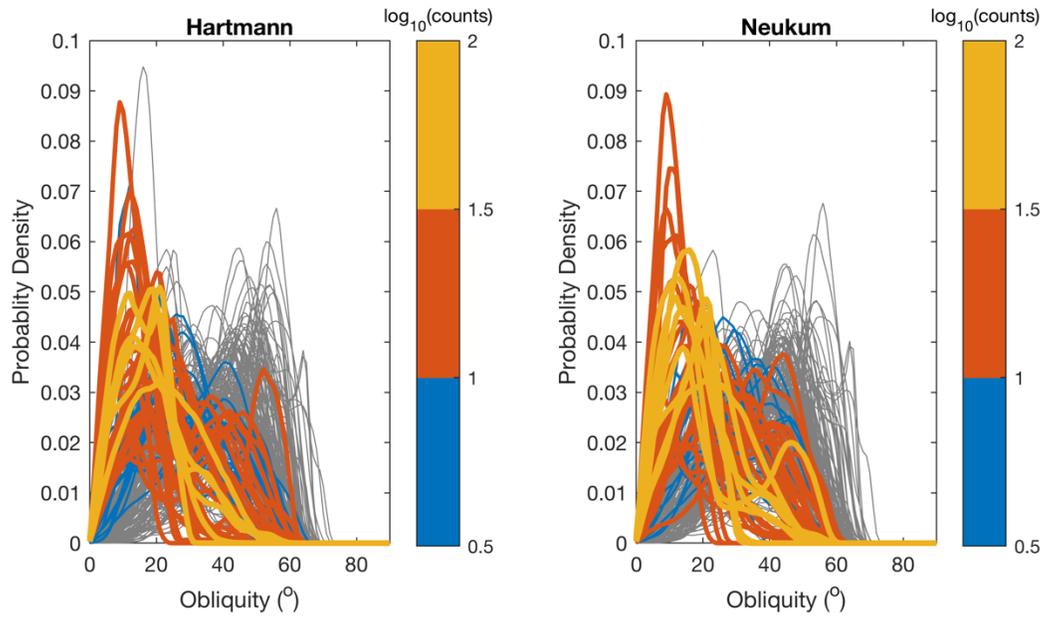

**Figure 8.** All 250 obliquity PDF's in our ensemble, each colored by the number of times it provides the best fit to a bootstrapped sample of the data (tracks selected fewer than 3 times are shown in grey).

**Note to Editor: High resolution file has been attached with submission. This figure has been formatted to be 2 column widths.**

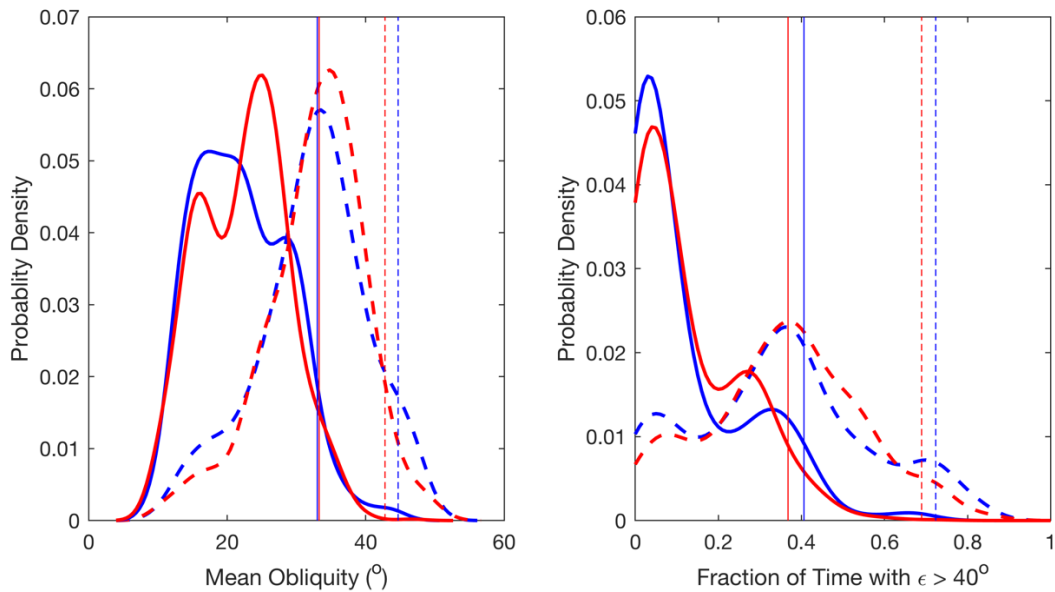

**Figure 9.** Gaussian kernel (bandwidth of 2° and 5% in the left and right plots respectively) smoothed PDF's of estimates of the mean obliquity (left) and fraction of the late Hesperian onward history spent at obliquity > 40 degrees. Hartmann case estimates are in blue while Neukum case are in red. The dashed lines represent the obliquity history ensemble prior to model application, and the solid lines represent the bootstrapped retrieved value. Vertical lines show 95$^{th}$ percentile locations.

**Note to Editor: High resolution file has been attached with submission. This figure has been formatted to be 2 column widths.**